Yutaka Takahashi

Department of Electrical and Information Engineering
Yamagata University
4-3-16 Jonan, Yonezawa-shi, Yamagata-ken 992
Japan
Tel. +81-238-26-3296(direct)
Fax +81-238-24-3240
e-mail  takahasy@eie.yz.yamagata-u.ac.jp

\documentstyle[prb,preprint,eqsecnum,aps]{revtex}
\begin{document}
\draft
\title{Monte Carlo Simulations of the Spatial Transport of
Excitons in a Quantum Well Structure}
\author{Yutaka~Takahashi\cite{email}}
\address{Department of Electrical and Information Engineering,
Yamagata University, Jonan, Yonezawa-shi, Yamagata 992, Japan}
\date{\today}

\maketitle

\begin{abstract}
The in-plane spatial transport of
nonequilibrium excitons in a GaAs quantum well structure
has been simulated with the ensemble Monte Carlo method.
The simulation has been performed for excitons
in the presence of residual heavy holes including the
interparticle Coulomb scatterings, LA phonon scatterings,
and exciton/carrier-interface roughness scatterings.
It has been found that, in contrast to the free electrons/holes
system in which the carrier-carrier scattering is significant,
the interface roughness scattering is
the dominant process for excitons because of
the relatively small scattering rate of
exciton-carrier and exciton-exciton scatterings.
This strongly affects
both the spatial motion and the energy relaxation of excitons.
The spatial and momentum distributions of excitons
have been simulated up to 500 ps at several exciton temperatures and
interface roughness parameters.
We have found that the exciton transport can be regarded as
a diffusive motion whose diffusion coefficient varies with time.
The diffusivity varies because the average velocity
of excitons change through the energy transfer between the
excitons and the residual heavy hole/lattice system.
\end{abstract}

\sloppy
\narrowtext
\section{Introduction}
In-plane transport of photogenerated electrons/holes
or excitons
confined in 2-dimensional (2D)
space of quantum well (QW) structures has
attracted much attention as it contains rich physics not accessible
in the stationary transport of conduction electrons.
It is also important from the practical point of view since the
spatial resolutions of position sensitive optical measurements such
as the photoluminescence (PL) via
optical microscopes and the recently-developed near-field scanning
optical microscopes (NSOM) \cite{Hess} are determined
by the diffusion length of photocarriers.

The measurements of photocarrier transport, which is a transient
phenomenon, are difficult
because it requires
high spatial and temporal resolutions.
There have been several attempts to measure it
in QW's of GaAs/AlGaAs with various
techniques under different experimental conditions.
The exciton localization was reported by Hegarty {\em et al.}
\cite{Hegarty} using the transient grating.
Oberhauser {\em et al.}\cite{Ober} discussed the scattering
mechanism of excitons with the same method.
The pump-probe technique was utilized by Smith {\em et al.}
\cite{Smith} and Yoon {\em et al.}\cite{Yoon} to investigate
the excitation density dependence of the spatial motion.
Tsen {\em et al.} \cite{Tsen1,Tsen2}
used the time and spatially resolved Raman
spectroscopy and discussed the spatial transport as well as
the energy relaxation.
The time-of-flight ({\em tof}) method
with a photomask was developed by Hillmer {\em et al.}
\cite{Hill1} and was applied to study the temperature and excitation
energy dependence of
the transport in QW's under low excitation conditions.
\cite{Hill2,Hill3,Tak1,Tak2}
The similar {\em tof} method with an optical fiber
was adopted by Akiyama {\em et al.},\cite{Akiyama}
who reported the transport properties in the wide range of excitation
density in doped and nondoped QW's.

In spite of these experimental
efforts the physical mechanism of transport is not
fully understood in the microscopic level so far.
This is because the spatial transport of photogenerated carriers
is closely related with the longitudinal relaxation of carriers
in the momentum space (i.e. the energy relaxation).
This makes the problem further complicated as compared with
the transport of conduction electrons for which the
momentum space distribution is stationary.
A distribution function of
photocarriers, which are initially generated
nonthermally at some point in the momentum space,
changes gradually due to
the scatterings of carriers, and
approaches the thermal equilibrium. At the same time carriers spread
spatially from the point where they are initially generated.
Since the scattering processes are affected both by the momentum
distribution and the spatial distribution (or the density)
of carriers and
the carrier velocities are determined by the momentum distribution,
the time evolutions of the momentum and the
spatial distribution are mutually dependent.
Thus it is necessary to solve the time evolution of momentum
and spatial distribution simultaneously to understand the
transport properties of photogenerated carriers.
Several authors\cite{Hill1,Basu}
reported the theoretical studies on the spatial
transport of excitons in QW's,
but they treated the momentum distribution
in thermal equilibrium, leaving its dynamic evolution out of account.

There are several approaches to the theoretical analysis of
the longitudinal momentum relaxation (energy relaxation)
of photogenerated carriers:\cite{Shah}
Analytical estimations based on Green's functions,
numerical integrations of Boltzmann
equations, and numerical simulations such as
ensemble Monte Carlo (EMC) method.
The last method is essentially the same as
solving (nonlinear) Boltzmann equations but,
with the advent of fast computers, it
has become widely used because it is possible to perform
quantitative calculations for nonequilibrium distributions
of carriers and easy to incorporate microscopic scattering
mechanisms into the model.
Another advantage of this method is that the realistic
experimental conditions such as the shape of the sample,
the initial, and the boundary conditions, can be used in the model
calculations, allowing one to compare the theory and the experiment
directly. (The details of this method is reviewed in Ref.
\onlinecite{Jacoboni}.)

There are many EMC studies on the
nonequilibrium dynamics of photogenerated carriers
\cite{Lugli1,Lugli2,Goodnick1,Artaki,Goodnick2,Lugli3,Bailey,Kuhn}
both in bulk and QW structures. But this method simulates
the dynamics of particles in momentum space and does not
yield the information in real space. Another class of numerical
simulation methods is the Molecular Dynamics (MD) approach, in which
all particle trajectories are calculated classically in real space.
Thus it naturally enables one to trace the motion of
particles in real space.
MD has been applied to the dynamics of nonequilibrium carriers
recently by several authors,\cite{Rota,Kane} however,
when the number of participating particles grows, it becomes
time-consuming and requires huge computer resources. An alternative
method is required to practically perform the numerical simulation
in the long time scale.

In this paper the spatial transport and the momentum relaxation
of excitons is numerically simulated.
The spatial motion is treated by dividing the simulation area by
mesh (concentric regions in the present case), and the
dynamics in the momentum space is calculated in each region
by EMC method at the
density and the momentum distribution averaged in the region.
The position and the momentum of the particle, and thus the
density and the momentum distribution in each region are updated
at every time step.
This approximation allows one to trace the spatial distribution
of particles at the spatial resolution comparable to the region size
without the expense of fast computational speed of EMC method.
The scattering mechanisms as carrier-carrier, exciton-exciton,
exciton-carrier, carrier-LA phonon, exciton-LA phonon,
carrier-interface roughness, and
exciton-interface roughness scatterings are
included in the simulations.

This paper is organized as follows:
In Sec. II the simulation conditions are briefly described,
followed by the detailed description the screening and scattering
mechanisms, and the method of EMC simulation.
In Sec. III {\em A} the comparison of free $e$/hh and exciton
relaxation is discussed and in
Sec. III {\em B} the numerical results of the exciton
spatial transport are given. They are discussed in comparison with
the previous experimental works. Conclusions are given in
Sec. IV.

\section{Model}
\subsection{Simulation Conditions}
The in-plane spatial motion and the energy relaxation of the
nonequilibrium excitons
are calculated in 2D space of nondoped QW structures of
GaAs, with the well width $L_z$ = 10 nm.
The infinite barrier height is assumed throughout.
The simulations have been performed in a circular area of
radius 10 $\mu$m.
The initial excitons are spatially distributed in the Gaussian shape,
$C\exp(-4\ln2 \frac{x^2}{FWHM^2})$,
with its full width at half maximum (FWHM) 6 $\mu$m,
and their momentum distribution is in the Bose distribution
(which is essentially identical to the Boltzmann distribution at
the density we have used), where the exciton temperature
$L_{ex}$ is used as a parameter for the initial momentum
distributions.
The peak areal density of excitons is
$1.9\times10^9$ cm$^{-2}$ at the center of the simulation area.
The residual carriers (heavy holes) of uniform areal density
$1\times10^9$ cm$^{-2}$
is assumed over the whole simulation area as with
Ref. \onlinecite{Goodnick1}, whose initial momentum distribution
is in the Fermi distribution whose temperature is the same as
the lattice at $T_L$.
($T_L \ne T_{ex}$)
In Sec. III {\em A} we also calculate the relaxation of free
$e$/hh's to compare with that of excitons. In this case we have
used mono-energetic initial momentum distributions both for
$e$/hh's and excitons.
The simulation is performed in the
exciton temperatures ($T_{ex}$) ranging between 10 and 90 K,
with the lattice temperature ($T_L$) and the interface roughness
parameters varied.
The electrons, heavy holes and excitons are treated in the parabolic
band approximation as it removes the complexity
of the expressions of carrier-carrier and
carrier-LA phonon scattering rates,
and contributes to the major
reduction of computational load.
Only the lowest subbands in the $\Gamma$ point of the conduction and
the heavy hole band is considered in the present study,
disregarding the contribution from the upper subbands and
the light hole bands.
Since the L-valley is located more than 280 meV above,
it does not contribute in the energy region of the present
simulation.
If we consider the realistic QW of Al$_{0.3}$Ga$_{0.7}$As/GaAs
with $L_z$ = 10 nm, the energy separations between
the lowest hh state and the second hh state or the first light
hole state is 22 meV and 13 meV, respectively.
(The separations are 33 meV and 48 meV, respectively, for
the well of the infinite barrier height.)
Since the average exciton energy at the highest exciton temperature
(90 K) is 7.7 meV, only a small fraction of excitons in the
high energy tail of the Bose distribution are involved in the
scatterings with the upper subbands. Thus we have expected that
the upper subbands does not appreciably
affect the relaxation and transport
properties, and we have not included them in the simulations.
The in-plane ($m_{hh//}$)
and perpendicular ($m_{hh\perp}$) (along the growth direction)
effective masses in the valence band are expressed using
Luttinger parameters, $\gamma_1$ and $\gamma_2$,
\cite{Luttinger}
$$m_{hh//}=\left( {\gamma _1 + \gamma _2} \right)^{-1}m_0,
\ \ \ \ m_{hh\perp}=\left( {\gamma _1 - 2\gamma _2} \right)^{-1}m_0$$
\noindent where $m_0$ is the free electron mass.
The physical parameters for GaAs in the present analysis
are taken from
Molenkamp {\em et al.};\cite{Molenkamp}
the effective mass in the conduction band $m_e=0.0665m_0$,
$\gamma_1=6.790$, and $\gamma_2=1.924$.

\subsection{Screening}
The proper treatment of the screening is essential in
the many-body theory of electrons
with electron-electron and electron-phonon
interactions. There are various approximated expressions
with the different levels of sophistication.\cite{Mahan,Haug1}
Among them the random phase approximation (RPA) is most commonly
used since it predicts important dynamical features such as
plasmons, and it has analytic expressions in 3D system both for the
degenerate\cite{Mahan} and
nondegenerate\cite{Meyer} carrier distributions.
But the analytic expression for 2D system is not obtained, and
we use the static expression in the present study.
The original RPA expression of the dielectric function,
which is valid in 2D for
any distributions of carriers,
is expressed as (see, e.g., Ref. \onlinecite{Haug1})

\begin{equation}
\label{RPA}
\varepsilon (\mbox{\boldmath $q$},\omega )=
1-V_q\sum\limits_{\mbox{\boldmath $k$,$\sigma$} }
{{{f_{\mbox{\boldmath $k$-$q$}}-f_{\mbox{\boldmath $k$}}} \over
{\hbar\omega +i\delta +E_{\mbox{\boldmath $k$-$q$}}
-E_{\mbox{\boldmath $k$}} }}},
\end{equation}

\noindent where $f_{\mbox{\boldmath $k$}}$ is the momentum
distribution function of carriers, $E_{\mbox{\boldmath $k$}}$ is
the energy of the carrier, and the summation runs over all
possible states of momentum and spin.
$V_q$ is the Fourier transform of the Coulomb potential
in 2D system:

\begin{equation}
\label{Vq}
V_q={{2\pi e^2} \over {\varepsilon _0L^2q}},
\end{equation}

\noindent where $L^2$ is the area of normalization,
and $\varepsilon _0$ is the dielectric constant including
the contributions from the interband electronic transitions
and from the phonons.
In the static ($\omega \to 0$) and the long wavelength limit
($\mbox{\boldmath $q$} \to 0$)
the RPA expression becomes,

\begin{equation}
\label{staticRPA}
\varepsilon (\mbox{\boldmath $q$}\to 0,\ 0)=
1-V_q\sum\limits_{k,\sigma }
{{{\partial f\left( {E_q} \right)} \over {\partial E_q}}}.
\end{equation}

\noindent For 2D, converting the summation over {\boldmath $k$} into
an integral over the energy $E$ by utilizing the parabolic
band dispersion,
assuming an isotropic momentum distribution,
and integrating by parts,
Eq. (\ref{staticRPA}) reduces to

\begin{equation}
\label{longstaticRPA}
\varepsilon (\mbox{\boldmath $q$}\to 0,0)=
1+\sum\limits_\sigma  {V_q{{L^2} \over {2\pi }}{{2m_e}
\over {\hbar ^2}}f\left( {E=0} \right)},
\end{equation}

\noindent which is valid for any isotropic
momentum distributions
of carriers. Using Eq. (\ref{Vq}) and summing up spin indices,
one finally obtains

\begin{equation}
\label{dielectricfunction}
\varepsilon (\mbox{\boldmath $q$}\to 0,0)=
1+{{2m_ee^2} \over {\varepsilon _0\hbar ^2q}}f\left( {E=0} \right)
=1+\frac{\kappa}{q},
\end{equation}

\noindent where the screening wavenumber $\kappa$ is defined by

\begin{equation}
\label{kappa}
\kappa \equiv {{2m_ee^2} \over {\varepsilon _0 \hbar ^2}}f
\left( {E=0} \right).
\end{equation}

\noindent
This expression for the dielectric function was first adopted by
Goodnick {\em et al.}\cite{Goodnick3}
and has been used for the EMC studies.\cite{Mosko}
We have used this expression throughout the present study.
This expression means that only the carriers in the
bottom of the band contribute to the screening.
The screened Coulomb interaction $V_q^{eff}$ is thus described as

\begin{equation}
\label{Coulomb}
V_q^{eff} \equiv  {{V_q} \over
{\varepsilon \left( {q\to 0,0} \right)}}\\
 = {{2\pi e^2} \over {\varepsilon _0 L^2}}
\left( {{1 \over {q+\kappa }}} \right),
\end{equation}

\noindent When the carriers from more than one band have to
be considered,
they contribute to the screening wavenumber independently in the
long wavelength limit ($q \to 0$).\cite{Haug2}
(Cross terms appear when $q$ is finite.)
Thus in the present case when the conduction and heavy hole band
take part in, the screening wavenumber is expressed as

\begin{equation}
\label{multibandkappa}
\kappa ={{2m_ee^2} \over {\varepsilon _0 \hbar ^2}}
\left[ {f_e\left( {E=0} \right)+f_{hh}\left( {E=0} \right)} \right]
\end{equation}

\noindent where $f_e\left( {E} \right)$
and $f_{hh}\left( {E} \right)$ are the momentum distribution
functions of electrons and heavy holes, respectively.
The screening wavenumber is calculated by counting
the number of electrons and heavy holes at the bottom of each band
in every time step of Monte Carlo simulations.
The strength of Coulomb interaction is updated every
time step using this screening wavenumber.

In order to obtain the full carrier dynamics in the relaxation,
the dynamic screening model is highly desirable.
However, the direct calculation of Eq. (\ref{RPA})
is computationally very heavy and beyond the reach of the
present study. (e.g. see Bair and Krusius in Ref. \onlinecite{Bair})
Instead, the static screening expression of Eq.
(\ref{dielectricfunction}) is used in the following simulations.

It should be noticed that the expressions for the screening is
valid only in the homogeneous system.
However, they can be applied to the inhomogeneous system
{\em locally}
when the characteristic length of the screening is much smaller
than the characteristic length at which the
distribution functions and the density change.
\cite{Hock}
In the present study the inverse of the screening wavenumber is
100 nm or less, while the characteristic length at which the exciton
distribution changes is of the order of 1 $\mu$m. Thus we can use
the locally defined screening in the present case.
Furthermore, as described
below, the excitons do not contribute to the screening but only the
residual heavy holes, whose distribution is almost homogeneous, are
responsible for the screening. Thus the effect of inhomogeneity
is further reduced.

We have not included excitons in the screening in the present
calculations. The screening in the presence of both charged
carriers and excitons are discussed in detail by Haug and
Schmitt-Rink\cite{Haug2} in the 3D system. In the simplest, case the
dielectric function due to excitons are given by
$$\varepsilon (\omega) = \varepsilon _0 (
1+ 4 \pi n \frac{9}{2} a_0^3),$$
where $n$ and $a_0$ are the exciton density and Bohr
radius, respectively.
In the present case this yields
$$\varepsilon = \varepsilon _0 (1+5.6 \times 10^{-2}).$$
On the other hand the dielectric function due to the free
carriers is
$$\varepsilon = \varepsilon _0 (1+ O(1)).$$
Thus the contribution from excitons is much smaller than that
from free carriers, and we have included only the free carriers
in the screening.

\subsection{Scattering Mechanism}
\label{scatmech}
\subsubsection{Carrier-Carrier Scattering}
The scattering between charged carriers is discussed in this
section.
The $e$-$e$, hh-hh, and $e$-hh scatterings in 2D using the
Born approximation has been discussed in
Ref. \onlinecite{Goodnick1} in detail and only the
results relevant to the present study are given here.
The total scattering rate of an electron of a momentum
$\mbox{\boldmath $k$}_1$ with heavy holes (momentum
$\mbox{\boldmath $k$}_2$) is given by

\begin{equation}
\label{ehtotalscat}
W_{e-h}(\mbox{\boldmath $k$}_1)=
{{2\pi e^4} \over {\varepsilon _0^2\hbar A}}{{m_r} \over {\hbar ^2}}
\sum\limits_{\mbox{\boldmath $k$}_2,\sigma }
{f_{hh}(\mbox{\boldmath $k$}_2,\sigma )}
\int_0^{2\pi } {d\theta {{\left| {F_{eehh}(q)} \right|^2}
\over {\left( {q+\kappa } \right)^2}}},
\end{equation}

\noindent where $m_r$ is the reduced mass, $q$ is the momentum
transfer from the electron to the heavy hole, and $\theta$ is the
scattering angle in c.m. frame.
$f_{hh}(\mbox{\boldmath $k$}_2,\sigma )$ is the
momentum distribution function of heavy holes.
The screening wavenumber $\kappa$ is given
by Eq. (\ref{dielectricfunction}).
$\mbox{\boldmath $q$}$ is related to
the initial and the final relative momentum,
$\mbox{\boldmath $k$}_r$ and $\mbox{\boldmath $k$}_r'$,
and the scattering angle $\theta$
of $e$-hh system by

\begin{equation}
\mbox{\boldmath $q$} =
\mbox{\boldmath $k$}_r - \mbox{\boldmath $k$}_r',
\; q = 2k_r\sin (\theta / 2)\nonumber
\end{equation}

\noindent where the relative momentum is defined by

\begin{displaymath}
\mbox{\boldmath $k$}_r=
{{m_{hh//}\mbox{\boldmath $k$}_1-m_{e//}\mbox{\boldmath $k$}_2}
\over {m_{hh//}+m_{e//}}}.
\end{displaymath}

\noindent $F_{eehh}(q)$ is the form factor given by

\begin{displaymath}
$$F_{eehh}(q)=\int\limits_{-\infty }^\infty
{dz_e\int\limits_{-\infty }^\infty
{dz_h}}\left| {\zeta _e(z_e)} \right|^2\left|
{\zeta _h(z_h)} \right|^2e^{-q\left| {z_e-z_h} \right|},
\end{displaymath}

\noindent where $\zeta _e(z_e)$ and $\zeta _h(z_h)$
are the envelope functions in the growth axis for electrons
and heavy holes, respectively. The form factor is
of the order of unity when
only the lowest subbands of the conduction and
the heavy hole bands are considered.
Since carrier densities are low in the nondegenerate region,
Pauli exclusion in the final state is not considered.
The similar expression holds for $e$-$e$ scattering
with antiparallel spins.
The exchange term is included when dealing with
$e$-$e$ scattering with parallel spins.
The total scattering rate in this case is given by
\cite{Mosko}

\begin{equation}
\label{eetotalscat}
W_{e-e}(\mbox{\boldmath $k$}_1)=
{{2\pi e^4} \over {\varepsilon _0^2\hbar A}}{{m_r} \over {\hbar ^2}}
\sum\limits_{\mbox{\boldmath $k$}_2}
{f_e(\mbox{\boldmath $k$}_2,\sigma )}
\int_{-{\pi  \over 2}}^{{\pi  \over 2}}
{d\theta \left| {{{F_{eeee}(q)} \over {q+\kappa }}-
{{F_{eeee}(Q)} \over {Q+\kappa }}} \right|}^2,
\end{equation}

\noindent where $Q$ is

\begin{displaymath}
Q = 2k_r\cos(\theta / 2).
\end{displaymath}

\noindent
The origin of the exchange effect is in the exclusion principle
between the identical particles.
When the two particles are identical, they cannot
come close enough. Thus the short range interaction does not
work between the identical particles.
The scale at which the particle indistinguishability
sets in can be of
the order of de Broglie wavelength.
In the present case, when the relative momentum
between two electrons is 0.6 nm$^{-1}$, the de Broglie wavelength
is 10 nm. We expect that the exchange effect becomes important when
the interaction range is equal to or less than the de Broglie
length.
The interaction range defined by the inverse of the
screening wavenumber is much longer in our case, around 100 nm.
Thus the exchange term does not affect much and the interaction
strength between the electrons with parallel spin is similar to
that between the electrons with antiparallel spin.

\subsubsection{Electron (Heavy Hole)-LA phonon Scattering}
The LA phonon scattering
is treated in the present study since
the energy of carriers and excitons is less than LO phonon
energy and the lattice temperature is low ($\le$ 30 K).
The 2D electrons interact with the phonons of bulk modes
through the deformation potential.
The scattering rate in 2D can be evaluated in parallel
with the exciton-LA
phonon interaction given by Takagahara in detail.\cite{Takagahara}
The total LA-phonon (wavenumber $Q_{ph}$) absorption and emission
rates from the electron with
\mbox{\boldmath $k$} are given by

\begin{eqnarray}
\label{LAscat}
W_{e-LA}^{absorb}(\mbox{\boldmath $k$}) &=&
{1 \over {4\pi }}{{D_e^2m_{e//}} \over {\hbar ^2u\rho L_z}}2k
\int_0^{2\pi } {d\theta {{\sin {\theta  \over 2}}
\over {e^{\hbar uQ_{ph}\// kT}-1}}}\\
W_{e-LA}^{emit}(\mbox{\boldmath $k$}) &=&
{1 \over {4\pi }}{{D_e^2m_{e//}} \over {\hbar ^2u\rho L_z}}2k
\int_0^{2\pi } {d\theta \sin {\theta  \over 2}
\left[ {{1 \over {e^{\hbar uQ_{ph}\// kT}-1}}+1} \right]},
\end{eqnarray}

\noindent where $u$, $\rho$, and $D_e$  are the velocity
of sound, the mass density, and the deformation potential for
conduction band, respectively.
$\theta$ is the electron scattering angle
in lab. frame.
The similar expression holds for heavy hole-LA phonon scattering.
Here we assume that the phonon energy
$\hbar uQ_{ph}$ is much smaller than the electron kinetic energy.
The z-component of the phonon momentum is
always set to zero for simplicity, resulting in the slight
overestimation of the scattering rate.
In the present analysis the physical parameters are
taken from Ref. \onlinecite{Takagahara};
$D_e$ = -6.5 eV, $D_{hh}$ = 3.1 eV, $\rho$ = 5.3 g cm$^{-3}$,
$u = 4.81 \times 10^5 $ cm s$^{-1}$.

In the present analysis the acoustic phonon distribution is
always assumed to be that of thermal equilibrium at the lattice
temperature, and the effects of nonequilibrium phonons emitted
from the carriers are not included since the carrier density and
the excitation energy is low.
There are several investigations,
theoretically and experimentally,
\cite{Smith,Zinov,Wolfe,Bulatov,Ramsbey} on the effects of
nonequilibrium phonons on the
dynamics of carriers and
excitons. This should be included in the case of high density, high
excitation energy.

\subsubsection{Electron (Heavy Hole)-Interface Roughness Scattering}
The electron (heavy hole)-interface roughness (IFR) scattering
in heterostructures
was given by Ando {\em et al.}\cite{Ando} and applied to
the transport phenomena of 2D electron gas in QW's.\cite{Gold,Sakaki}
Here the brief derivation is given for clarity.
In the well of infinite barriers, the lowest subband energy as a
function of well width is given by
$
E(L_z)=\pi ^2\hbar ^2 / (2m_{e\perp}L_z^2).
$
When the well width changes from place to place
in the 2D plane by
$\Delta L_z(\mbox{\boldmath $r$}_{//})$,
the subband energy changes by

\begin{equation}
\label{deltaE}
\Delta E(\mbox{\boldmath $r$}_{//})=
{{\pi ^2\hbar ^2\Delta L_z(\mbox{\boldmath $r$}_{//})}
\over {m_{e\perp}L_z^3}},
\end{equation}

\noindent where $\mbox{\boldmath $r$}_{//}$ is
the coordinate in 2D plane and $m_{e\perp}$ is the effective mass
perpendicular to 2D plane.
The fluctuation of the well width can be expanded by Fourier
series as

\begin{equation}
\label{fourierexp}
\Delta L_z(\mbox{\boldmath $r$}_{//})=
\sum\limits_{\mbox{\boldmath $q$}_{//}}
{\Delta _{\mbox{\boldmath $q$}_{//}}
e^{i\mbox{\boldmath $q$}_{//}
\cdot \mbox{\boldmath $r$}_{//}}}.
\end{equation}

\noindent $\Delta E(\mbox{\boldmath $r$}_{//})$ in Eq. (\ref{deltaE})
can be regarded as
a potential for 2D electrons and the $e$-IFR scattering amplitude
is given by its matrix element between 2D
plane wave states as

\begin{equation}
\label{eIRS}
\left\langle {\mbox{\boldmath $k$}'_{//}}
\right|\Delta E(\mbox{\boldmath $r$}_{//})
\left| \mbox{\boldmath $k$}_{//} \right\rangle =
{{\pi ^2\hbar ^2} \over {m_{e\perp}L_z^3}}
\Delta _{\mbox{\boldmath $q$}_{//}}.
\end{equation}

\noindent where $\mbox{\boldmath $k$}_{//}
-\mbox{\boldmath $k$}'_{//}=\mbox{\boldmath $q$}_{//}.$
The Gaussian correlation function of
the well-width fluctuation is assumed, thus

\begin{equation}
\label{gaussian}
\left\langle {\Delta L_z(\mbox{\boldmath $r$}_{\// /})
\cdot \Delta L_z(\mbox{\boldmath $r$}'_{\// /})} \right\rangle
=\Delta ^2\exp \left[ {\left( {\mbox{\boldmath $r$}_{//}
-\mbox{\boldmath $r$}'_{//}} \right)^2\/ /\Lambda ^2} \right],
\end{equation}

\noindent where $\Delta$ and $\Lambda$ are the amplitude and
the correlation length
of the fluctuation, respectively. Using the Fourier expansion
Eq. (\ref{fourierexp}), the left side reduces to

\begin{displaymath}
\left\langle {\sum\limits_{\mbox{\boldmath $q$}_{//},
\mbox{\boldmath $q$}'_{//}}
{\Delta _{\mbox{\boldmath $q$}_{\// /}}
\Delta _{\mbox{\boldmath $q$}'_{\// /}}
e^{i(\mbox{\boldmath $q$}_{//}\cdot \mbox{\boldmath $r$}_{//}
+\mbox{\boldmath $q$}'_{//}\cdot \mbox{\boldmath $r$}'_{//})}}}
\right\rangle
=
\sum\limits_{\mbox{\boldmath $q$}_{//}}
{\left| {\Delta _{\mbox{\boldmath $q$}_{\// /}}}
\right|}^2e^{i\mbox{\boldmath $q$}_{//}
\cdot \Delta \mbox{\boldmath $r$}_{//}}
\end{displaymath}

\noindent The right side can be expanded in Fourier series, then
Eq. (\ref{gaussian}) reduces to

\begin{displaymath}
\sum\limits_{\mbox{\boldmath $q$}_{//}}
{\left| {\Delta _{\mbox{\boldmath $q$}_{\// /}}}
\right|}^2e^{i\mbox{\boldmath $q$}_{//}
\cdot \Delta \mbox{\boldmath $r$}_{//}}=
\sum\limits_{\mbox{\boldmath $q$}_{\// /}}
{{1 \over A}\pi \Lambda ^2\Delta ^2e}^{-q_{//}^2\Lambda ^2\/ 4}
e^{i{\mbox{\boldmath $q$}_{\// /}}
\cdot \Delta {\mbox{\boldmath $r$}_{\// /}}}.
\end{displaymath}

\noindent Thus the expansion coefficient
is expressed as
$${\left| {\Delta _{\mbox{\boldmath $q$}_{\// /}}} \right|}^2=
{1 \over A}\pi \Lambda ^2\Delta ^2e^{-q_{//}^2\Lambda ^2\/ /4}.$$
Using this in the right side of Eq. (\ref{eIRS}), the square of
matrix element is written as

\begin{equation}
\label{eIRSmatrix}
\left| {\left\langle {\mbox{\boldmath $k$}'_{\// /}}
\right|\Delta E(\mbox{\boldmath $r$}_{\// /})
\left| {\mbox{\boldmath $k$}_{\// /}}  \right\rangle} \right|^2
={{\pi ^5\hbar ^4\Lambda ^2\Delta ^2} \over {Am_{e\perp}^2L_z^6}}
e^{-q_{//} ^2\Lambda ^2\/ /4}.
\end{equation}

\noindent The total scattering rate of the electron with momentum
$\mbox{\boldmath $k$}$ is calculated via Fermi's golden rule as

\begin{equation}
\label{eIRSrate}
W_{e-IFR}({\mbox{\boldmath $k$}_{\// /}})=
{{\pi ^4\hbar \Lambda ^2\Delta ^2m_{e//}}
\over {2m_{e\perp}^2L_z^6}}\int\limits_0^{2\pi } {d\theta }
e^{-k_{//}^2\Lambda ^2\sin ^2\left( {\theta \/ /2} \right)}
\end{equation}

\noindent where $\theta$ is the scattering angle in Lab. frame.
The experimental determination of IFR parameters,
$\Delta$ and $\Lambda$, is very difficult. The former is
the well width fluctuation and usually one or two monolayers
is assumed. The latter corresponds to the typical terrace or
island size in the two dimensional plane.
In the present simulations $\Delta$ is 0.283 nm (one monolayer)
and $\Lambda$ is 10 nm unless otherwise stated.
In this model, since the infinite barrier height is assumed,
the IFR scattering is somewhat overestimated compared with
the realistic well of a finite barrier height.

\subsubsection{Exciton-Electron (Heavy Hole) Scattering}
\label{sect-e-exscat}
We derive here the exciton-electron (ex-$e$) scattering to the
lowest order.
The exciton wavefunction in the ideal 2D system (z-dependence
ignored) is written as
$$\left| \mbox{\boldmath $K$} \right\rangle={1 \over {\sqrt A}}\sum
\limits_{\mbox{\boldmath $r$}_e,{\mbox{\boldmath $r$}_h}}
{e^{i\mbox{\boldmath $K$}\cdot \mbox{\boldmath $R$}}
F(\mbox{\boldmath $r$}_e,\mbox{\boldmath $r$}_h)
c_{\mbox{\boldmath $r$}_e}^{\dag}b_{\mbox{\boldmath $r$}_h}^{\dag}
\left| 0 \right\rangle},$$
where $\mbox{\boldmath $K$}$ and $\mbox{\boldmath $R$}$
are the center-of-mass momentum and coordinate, respectively.
$F$ is the wavefunction of relative motion, and
$c_{\mbox{\boldmath $r$}_e}^{\dag}$
($b_{\mbox{\boldmath $r$}_h}^{\dag}$) is the electron (heavy hole)
creation operator in the Wannier representation.
The wavefunction can be rewritten in the Bloch representation
using the relation
$$c_{\mbox{\boldmath $r$}_e}^{\dag}=
{1 \over {\sqrt N}}\sum\limits_{\mbox{\boldmath $k$}}
{e^{-i\mbox{\boldmath $k$}\cdot \mbox{\boldmath $r$}_e}
c_{\mbox{\boldmath $k$}}^{\dag}},$$
where $N$ is the number of unit cells in the area $A$,
$A=Nv_0$ ($v_0$ is the area of the unit cell).
Converting the summation over lattice sites into the integral by
$$\sum\limits_{\mbox{\boldmath $r$}} {}\to {1 \over {v_0}}
\int {d^2\mbox{\boldmath $r$}},$$
the exciton wavefunction reduces to

\begin{equation}
\label{exwavefunc}
\left| \mbox{\boldmath $K$} \right\rangle=\sum
\limits_{\mbox{\boldmath $k$},\mbox{\boldmath $k$}'}
{f(\mbox{\boldmath $k$},\mbox{\boldmath $k$}',\mbox{\boldmath $K$})
\delta _{\mbox{\boldmath $K$}, \mbox{\boldmath $k$}
+\mbox{\boldmath $k$}'} c_{\mbox{\boldmath $k$}}^{\dag}
b_{\mbox{\boldmath $k$}'}^{\dag}\left| 0 \right\rangle},
\end{equation}

\noindent where

\begin{equation}
\label{ftoF}
f(\mbox{\boldmath $k$},\mbox{\boldmath $k$}'
,\mbox{\boldmath $K$})={N \over {A^{3\/ /2}}}\int {d^2
\mbox{\boldmath $r$}F(\mbox{\boldmath $r$})
e^{i(\alpha _e\mbox{\boldmath $K$}-\mbox{\boldmath $k$})
\cdot \mbox{\boldmath $r$}}}.
\end{equation}

\noindent The mass ratios $\alpha _e$ and $\alpha _h$ are defined by
$$\alpha _e={{m_e} \over {m_e+m_h}},
\quad\alpha _h={{m_h} \over {m_e+m_h}}.$$

We evaluate ex-$e$ (or ex-hh)
scattering to the lowest order,
assuming that there happens neither rearrangement of electrons
nor the excitations of the internal state of the exciton,
and ignoring exchange effect.
The Coulomb interaction Hamiltonian is
written as (spin indices are dropped)
$$H'={1 \over 2}\sum\limits_{\mbox{\boldmath $k$},
\mbox{\boldmath $k$}',\mbox{\boldmath  $q$}\ne 0}
{V_q^{eff}\alpha _{\mbox{\boldmath $k$}+\mbox{\boldmath $q$}}^{\dag}
\alpha _{\mbox{\boldmath $k$}'-\mbox{\boldmath $q$}}
^{\dag}\alpha _{\mbox{\boldmath $k$}'}}
\alpha _{\mbox{\boldmath $k$}},$$
where $\alpha _{\mbox{\boldmath $k$}}^{\dag}$ is
an operator for electrons or heavy holes. $V_q^{eff}$ is
the screened Coulomb potential of Eq. (\ref{Coulomb}),
where the screening in only due to
the free $e$/hh's.
The scattering amplitude is obtained by calculating
the matrix element of $H'$ between the initial and the final
electron-exciton states,
$\left| \mbox{\boldmath $k$},\mbox{\boldmath $K$} \right\rangle$ and
$\left| \mbox{\boldmath $k$}',\mbox{\boldmath $K$}' \right\rangle.$
Here $\mbox{\boldmath $k$}$ ($\mbox{\boldmath $k$}'$) and
$\mbox{\boldmath $K$}$ ($\mbox{\boldmath $K$}'$) represent
the initial (final) electron and exciton momentum, respectively.
Two terms remain and the corresponding diagrams are
shown in (a) and (b) of Fig. \ref{diagrams}.
The term described by (a) is

\begin{displaymath}
\sum\limits_{\mbox{\boldmath $k$}_1,\mbox{\boldmath $k$}_2,
\mbox{\boldmath $l$}_1,\mbox{\boldmath $l$}_2}
{V_qf^{\ast}(\mbox{\boldmath $l$}_1,\mbox{\boldmath $l$}_2,
\mbox{\boldmath $K$}')f(\mbox{\boldmath $k$}_1,
\mbox{\boldmath $k$}_2,\mbox{\boldmath $K$})
\delta _{\mbox{\boldmath $K$}',\mbox{\boldmath $l$}_1+
\mbox{\boldmath $l$}_2}}\delta _{\mbox{\boldmath $K$},
\mbox{\boldmath $k$}_1+\mbox{\boldmath $k$}_2}
\delta _{\mbox{\boldmath $q$},\mbox{\boldmath $l$}_1-
\mbox{\boldmath $k$}_1}\delta _{\mbox{\boldmath $l$}_2,
\mbox{\boldmath $k$}_2}.
\end{displaymath}

\noindent This term can be evaluated
by using Eq. (\ref{ftoF}) and converting the summation over
momenta into the integral. Evaluating two terms (a) and (b),
the matrix element is give by

\begin{equation}
\label{eexmatrix1}
\left\langle {\mbox{\boldmath $k$}_3-\mbox{\boldmath $q$},
\mbox{\boldmath $K$}+\mbox{\boldmath $q$}} \right|H'
\left| {\mbox{\boldmath $k$}_3,\mbox{\boldmath $K$}} \right\rangle
={{V_qN^2} \over {A^2}}\int {d^2\mbox{\boldmath $r$}
\left| {F(r)} \right|^2
\left[e^{i\alpha _h\mbox{\boldmath $q$}
\cdot \mbox{\boldmath $r$}} -
e^{-i\alpha _e\mbox{\boldmath $q$}
\cdot \mbox{\boldmath $r$}}\right]}.
\end{equation}

\noindent Using the internal wavefunction of
the lowest energy, $s$-wave exciton;
$F(r)=2\alpha v_0 e^{-\alpha r}/ \sqrt {2\pi},$
where $\alpha$ is the inverse of the Bohr radius,
Eq. (\ref{eexmatrix1}) reduces to

\begin{equation}
\label{eexmatrix2}
{{V_qN^2} \over {A^2}}{{4\alpha ^2v_0^2} \over {2\pi }}
\int {rdrd\theta e^{-2\alpha r}\left[ {e^{i\alpha _h
\mbox{\boldmath $q$}\cdot \mbox{\boldmath $r$}}
-e^{-i\alpha _e\mbox{\boldmath $q$}\cdot
\mbox{\boldmath $r$}}} \right]}.
\end{equation}

\noindent This expression can be integrated by
using the generation function of the Bessel function,
and one finally obtains for the matrix element

\begin{equation}
\label{eexmatrix3}
\left\langle {\mbox{\boldmath $k$}_3-\mbox{\boldmath $q$},
\mbox{\boldmath $K$}+\mbox{\boldmath $q$}} \right|H'\left|
{\mbox{\boldmath $k$}_3,\mbox{\boldmath $K$}} \right\rangle=
V_q\left\{ {{1 \over {\left[ {1+\left( {{{\alpha _hq}
\over {2\alpha }}} \right)^2} \right]^{3\/ /2}}}-
{1 \over {\left[ {1+\left( {{{\alpha _eq}
\over {2\alpha }}} \right)^2} \right]^{3\/ /2}}}} \right\}.
\end{equation}

\noindent The total scattering rate of an exciton with electrons
(heavy holes)
can be calculated with the Fermi's golden rule as

\begin{equation}
\label{eexscatrate}
W_{ex-e,h}(\mbox{\boldmath $K$})={{2\pi e^4m_r}
\over {\hbar ^3\varepsilon _0^2A}}\sum\limits_{\mbox{\boldmath $k$}}
{f_{e,h}(\mbox{\boldmath $k$})\int_0^{2\pi } {d\theta
{1 \over {\left( {q+\kappa } \right)^2}}\left\{ {{1
\over {\left[ {1+\left( {{{\alpha _hq} \over {2\alpha }}}
\right)^2} \right]^{3\/ /2}}}-{1 \over {\left[ {1+
\left( {{{\alpha _eq} \over {2\alpha }}} \right)^2} \right]
^{3\/ /2}}}} \right\}^2}},
\end{equation}

\noindent where $m_r$ is the reduced mass of an exciton and electron
(heavy hole), $f_{e,h}(\mbox{\boldmath $k$})$ is the momentum
distribution function of electrons (heavy holes), and $\kappa $
is the  screening wavenumber defined in Eq. (\ref {kappa}).
The scattering angle $\theta$ is related to the momentum transfer
$q$ and the relative momentum $m_r$ as
$q=2k_r/\sin (\theta /2).$
The exciton Bohr radius $\alpha^{-1}$
is estimated to be 12.5 nm from
the variational calculation.\cite{Tak3}

\subsubsection{Exciton-Exciton Scattering}
We can derive the exciton-exciton (ex-ex)
scattering in the similar manner
as ex-$e$ scattering described in the previous section.
The relevant diagrams are shown in (c)-(f) of Fig. \ref{diagrams}.
The total scattering rate is given as

\begin{equation}
\label{exexscatrate}
W_{ex-ex}(\mbox{\boldmath $K$}_{\alpha})=16{{2\pi e^4m_r}
\over {\hbar ^3\varepsilon _0^2A}}
\sum\limits_{\mbox{\boldmath $K$}_{\beta}}
{f_{ex}(\mbox{\boldmath $K$}_{\beta})\int_0^{2\pi } {d\theta
{1 \over {\left( {q+\kappa } \right)^2}}\left\{ {{1
\over {\left[ {1+\left( {{{\alpha _hq} \over {2\alpha }}}
\right)^2} \right]^{3\/ /2}}}-{1 \over {\left[ {1+
\left( {{{\alpha _eq} \over {2\alpha }}} \right)^2} \right]
^{3\/ /2}}}} \right\}^4}},
\end{equation}

\noindent where $\mbox{\boldmath $K$}_{\alpha}$ is the momentum
of the exciton, $m_r$ is the reduced mass of two excitons, and
$f_{ex}(K)$ is the momentum distribution function of excitons.

Comparing the exciton scattering rates, Eq. (\ref{eexscatrate}) and
Eq. (\ref{exexscatrate}), with the electron scattering rate,
Eq. (\ref{ehtotalscat}), they show large differences in their
magnitude and $q$-dependence. Figure \ref{compscatrate}
shows the plots of the scattering rates of the three
processes vs. $q$.
The $e$-hh scattering rate is forward peaked, and
is larger than other two processes by several orders of
magnitude. The magnitudes of
ex-hh and ex-ex scattering
rise with $q$ and reach the peak at $q=0.15$ and $0.20$ nm$^{-1}$,
respectively, and are much smaller than $e$-hh scattering.
These features of scattering mechanism lead to the
striking difference in the relaxation and transport between electrons
and excitons as described later.

\subsubsection{Exciton-LA Phonon Scattering}
There is a detailed theoretical study of Takagahara\cite{Takagahara}
on the exciton-LA phonon interaction and only the results
relevant to the present calculations are shown here.
The total exciton scattering rates for the phonon absorption
and emission processes have the similar forms as the electron case
as,

\begin{eqnarray}
\label{exLAscat}
W_{ex-LA}^{absorb}(\mbox{\boldmath $k$}) &=&
{1 \over {4\pi }}{{(D_e-D_{hh})^2m_{e//}} \over
{\hbar ^2u\rho L_z}}2k
\int_0^{2\pi } {d\theta {{\sin {\theta  \over 2}}
\over {e^{\hbar uQ_{ph}\// kT}-1}}}\\
W_{ex-LA}^{emit}(\mbox{\boldmath $k$}) &=&
{1 \over {4\pi }}{{(D_e-D_{hh})^2m_{e//}} \over
{\hbar ^2u\rho L_z}}2k
\int_0^{2\pi } {d\theta \sin {\theta  \over 2}
\left[ {{1 \over {e^{\hbar uQ_{ph}\// kT}-1}}+1} \right]},
\end{eqnarray}

\noindent where the notations are the same as electron-LA phonon
scattering.

\subsubsection{Exciton-Interface Roughness Scattering}
The exciton-IFR scattering is deduced from
the electron-IFR scattering.
Using the matrix element of Eq. (\ref{eIRSmatrix}),
the electron-IFR interaction Hamiltonian can be written as

\begin{equation}
\label{HeIR}
H_{ex-IFR}'=\sum\limits_{\mbox{\boldmath $k$}_{//},
\mbox{\boldmath $q$}_{//}}
{{\pi ^{5/2}\hbar ^2\Lambda \Delta } \over {A^{1/2}L_z^3}}
e^{-q_{//} ^2\Lambda ^2\/ /2}
{({ 1\over {m_{e\perp}}}c_{\mbox{\boldmath $k$}_{//}
+\mbox{\boldmath $q$}_{//}}^{\dag}c_{\mbox{\boldmath $k$}_{//}}+
{ 1\over {m_{hh\perp}}}b_{\mbox{\boldmath $k$}_{//}
+\mbox{\boldmath $q$}_{//}}^{\dag}b_{\mbox{\boldmath $k$}_{//}})}
\end{equation}

\noindent where $c^{\dag}$ and $b^{\dag}$ are operators for
an electron and a heavy hole, respectively.
The derivation is similar to the case of the $e$-IFR
scattering and
the total scattering rate is given by

\begin{equation}
\label{exIFR}
W_{ex-IFR}(\mbox{\boldmath $K$})=
{{\pi ^4\hbar \Delta ^2\Lambda ^2m_{ex//}}
\over {2L_z^6}}\left( {{1 \over {m_{e\perp }}}-{1
\over {m_{hh\perp }}}} \right)^2\int\limits_0^{2\pi }
{d\theta e^{-K^2_{//}\Lambda ^2\sin ^2{\theta \over {2}}}}.
\end{equation}

\subsection{Simulation Method}
\label{simmethod}
The EMC simulations in $k$-space have been
widely used in the analysis of photogenerated carriers
\cite{Lugli1,Lugli2,Goodnick1,Artaki,Goodnick2,Lugli3,Bailey,Kuhn}
and reviewed in Ref. \onlinecite{Shah}, and the detailed technique
is not repeated here.
In order to trace the distributions of the particles
in $r$-space within the frame work of the $k$-space EMC method,
the simulation area is divided
into ten concentric circular regions as shown in Fig. \ref{simarea}.
In each region the average particle density and the momentum
distribution function are calculated in every
time step and the $k$-space EMC simulation is performed,
independently from other regions, with the scattering
parameters corresponding to the average density.
(Interparticle scatterings are affected directly through
their dependence on the density and indirectly through
the change in the screening wavenumber. Other scatterings are
also affected indirectly through the change of the momentum
distribution of particles.)
The positions, as
well as the momenta, of particles
are updated after each time step.
A particle can move to the neighboring region, thus the particle
density, and consequently the scattering parameters, change
every time step. This technique allows one to trace the time
evolution of both the spatial and the momentum
distribution of particles with relatively low computational load.
The disadvantage is that, since the continuous
density distribution (solid line in Fig. \ref{simarea}) is
approximated by the step-like one, the difference of the
scattering rates between the particles near the inner
boundary and the outer boundary
of the region is neglected. Thus the spatial distribution of
particles calculated in the present method is only valid in
the length scale larger than the region size.
The range of the Coulomb interaction, determined by the
inverse of the screening wavenumber is of the order of
100 nm in the present simulations,
which is much smaller than the region size.

In the present study ten or twenty simulation areas as
mentioned above are prepared and simulated simultaneously.
The physical information is obtained by taking their
ensemble average.
The number of particles is 78600 when twenty simulation areas
are used.
The time step of simulations are 5 fs for free $e$/hh's
and 12.5 fs for excitons. We have confirmed that
the time step is much smaller than the scattering interval and that
the simulation results do not change by further reducing the time
step.

\section{Results and Discussion}
\subsection{Electron and Exciton Relaxations}
It is expected that
the scattering frequency and thus the energy relaxation is different
between free $e$/hh's and excitons reflecting the different
magnitude of scattering processes.
To see this, the frequency of each scattering process is counted
from 0 to 15 ps both for electrons and excitons.
For $e$/hh's the simulation is done in the mono-energetic initial
momentum distribution,
with the excitation energy 10 meV above the band gap
(the excess energies is 6.34 meV for electrons and 3.66 meV for
heavy holes) in the presence of residual hh's
at the lattice temperature 5 K.
The spatial distribution is in the Gaussian function with FWHM
6 $\mu$m as with excitons.
The similar simulation is performed for excitons assuming that
they were generated mono-energetically at 10 meV.
Notice that this is physically not realistic because it is not
possible to optically generate excitons except at the bottom of
the band due to the energy-momentum conservation.
It is calculated to show the remarkable difference between the
electron and exciton relaxation due to the difference in the
magnitude of interparticle scatterings.

Plotted in Fig. \ref{No.ofScat} are the numbers of each
scattering events per particle per second.
For electrons $e$-hh, $e$-$e$ and
$e$-IFR scatterings are of the similar magnitude of
$1\times10^{12}$ s$^{-1}$.
(The frequency the $e$-LA phonon scatterings is
several orders of magnitude smaller than those plotted here.)
The $e$-hh scattering frequency increases in the initial few
ps reflecting the rapid energy relaxation of electrons:
When the electrons
have larger momenta and there are many residual hh's with
very small average momentum just after photogeneration,
$e$-hh scattering is less frequent because it favors
the small momentum transfer.

The exciton scatterings plotted in the lower part of
Fig. \ref{No.ofScat} show different features.
(Notice that they are plotted in logarithmic scale.)
While the ex-IFR scattering is as frequent as
$e$-IFR scattering,
the interparticle scatterings as ex-ex and ex-hh scatterings
are more than two orders of magnitude smaller than
those of electrons.
This is due to the small scattering rates of excitons
as shown in Fig. \ref{compscatrate}.
Thus the contributions from the interparticle scatterings
(and LA phonon scatterings which is
not shown in  Fig. \ref{No.ofScat} as the figure becomes
too busy) is very small.
The dominance of IFR scattering in excitons is in agreement
with the analytic calculations of Basu {\em et al.} \cite{Basu}

The difference of the interparticle scattering rates
in free $e$/hh and
exciton system has significant effect on the time evolutions of
the momentum distribution functions.
Figure \ref{energyrelaxation} shows the distribution functions
of the initial 15 ps. The electrons, initially generated at
0.105 nm$^{-1}$ (= 6.337 meV), relax rapidly, and approach the
thermal distribution described by the Boltzmann distribution
function (with the electron temperature $T_e$) at 15 ps.
This is due to the high rate of $e$-hh and $e$-$e$ scatterings
which are quite efficient in redistributing kinetic energies
among electrons and heavy holes. It should be noticed that
both the electrons and the heavy holes are in equilibrium, with the
same temperature $T_e$. (The thermalization with the lattice
is achieved through the interaction with LA-phonons. But this
process is quite slow, more than two orders of magnitude lower
than $e$-$e$ scattering. Thus the electron temperature is
larger than the lattice temperature at this stage.)
On the other hand the exciton distribution does not change
appreciably in the time scale shown here. The ex-IFR scattering,
which is the dominant process for excitons,
is an elastic scattering, and does not change the kinetic energy
(or the magnitude of momentum) of excitons.
Thus the energy relaxation is induced only through very slow
ex-hh and LA phonon scatterings.
Another feature attributed to the difference of the
scattering processes is that the electron transport would
depend on the density while the exciton transport would not.

\subsection{Exciton Transport for 500 ps}
The spatial transport and the energy relaxation of excitons have
been calculated up to 500 ps. Here the initial excitons are
in the Bose distribution with the temperature
$T_{ex}$ while the initial residual heavy holes are in the
Fermi distribution at $T_L$ in equilibrium with the lattice.
Figure \ref{excitonrelax} shows the time evolution in
the momentum distribution functions of excitons
up to 500 ps with 25 ps step, at $T_{ex}=$ 30 K and the lattice
temperature is 5 K.
Because of the low scattering rates of ex-ex,
ex-hh, and ex-LA phonon processes, the energy relaxation
is slow, and the excitons do not reach the thermal equilibrium with
the lattice even at 500 ps.
Thus the analytic calculations of the exciton transport
which assume the momentum distribution in equilibrium
with the lattice can underestimate the diffusivity
in the earlier stage.

We have obtained the full width at half maximum (FWHM) of the
exciton spatial distribution by fitting the Gauss function to
the distribution calculated from the simulations.
Figure \ref{Fig7FWHM} shows the time evolution of
the spatial spread (FWHM) of
excitons at several initial exciton temperatures ($T_{ex} =$
10 K - 90 K) from 0 ps to 500 ps when the initial residual
hh's (and the lattice) are at 5 K.
The fitting uncertainty in FWHM is around 0.03 $\mu$m.
The calculations show that the excitons spread faster
when the initial exciton temperature is higher.
This trend can be easily understood when we remember that the
dominant scattering process for excitons is the ex-IFR scattering.
The ex-IFR scattering probability (Eq. (\ref{exIFR})) does not
depend on the exciton density and is weakly dependent on the
exciton momentum. Thus the the mean free path
of excitons due to the ex-IFR scattering is similar at any
place in the simulation area and does not vary significantly
during the simulation from 0 to 500 ps.
In this case, the transport is essentially
determined by the velocities of excitons, with which
they travel during consecutive IFR scatterings.
Since the average velocity is directly related to the exciton
temperature, the transport becomes enhanced
with the temperatures.
This also means that, in the short time scale where the exciton
temperature is virtually constant,
the transport can be regarded as diffusive
motion characterized by the constant diffusion coefficient,
$D=\langle v \rangle \lambda /2,$ over whole simulation area.
Here $\langle v \rangle$ is an average velocity and
$\lambda$ is a mean free path.
If excitons were to  go through diffusive motion with a constant
diffusion coefficient and the initial spatial distribution were
Gaussian, they would spread with time $t$ retaining
Gaussian-shaped distribution
with its $FWHM(t)$ given by $\sqrt{16\ln2 \times Dt+FWHM(0)^2}$.
We have also plotted this curve with the diffusion constant
30 cm$^2$ s$^{-1}$ in Fig. \ref{Fig7FWHM} with a broken line
for reference.
The FWHM with a constant diffusion coefficient is almost linear in
the scale of Fig. \ref{Fig7FWHM}, in contrast to the sublinear
trend of the simulation results.
This can be interpreted that the diffusion coefficient decreases
dynamically due to the reduction of the average velocity.
This is because
the initial exciton temperatures are higher than the initial
temperature of residual hh's or the lattice, the exciton temperature
(and their average velocity) decreases through the ex-hh and
ex-LA phonon interactions. The diffusion coefficient of the initial
stage (0 - 25 ps) is 180 cm$^2$ s$^{-1}$ for $T_{ex}=$ 90 K, while
it reduces to 5 cm$^2$ s$^{-1}$ at 475 - 500 ps.

In Fig. \ref{Fig8FWHM} we have plotted the spatial spread of
excitons when $T_L$ (the initial residual hh temperature and the
lattice temperature) is varied. In the upper plot, when
$T_{ex}$ (30 K) is larger than $T_L$ (5 K) (solid curve),
the FWHM is sublinear, reflecting the cooling down of excitons
by the residual hh's and the lattice.
When $T_{ex} = T_L =$ 30 K (broken curve),
the FWHM increases linearly because there is no net energy flow
among excitons, residual hh's and the lattice.
Actually in this case the FWHM evolution can
be reproduced by the simple diffusion model with a constant
diffusivity $D=$ 19 cm$^2$ s$^{-1}$.
On the other hand, in the lower plot,
when the lattice temperature (30 K) is higher
than the initial exciton temperature (10 K) (broken curve),
the FWHM evolution is superlinear because the excitons are
heated by the residual hh's and the lattice.

In the calculations above, the IFR parameters were fixed
($\Delta= 0.283$ nm, $\Lambda = 10$ nm). We have done further
calculations to investigate how the interface
roughness affects the transport phenomena.
In the present model the interface roughness is characterized
by the two parameters; $\Lambda$ and $\Delta$.
The former designates the typical lateral size of terrace or island
in the 2D plane of QW's, while the latter indicates the terrace
height.
There are several experimental investigations to observe
the interface structures in QW's
\cite{Sakaki,Christen,Zachau,Gust}, but it is still a
controversial problem.
We have used several typical values of the parameters in the
simulations.
Figure \ref{Fig9IFR} shows the FWHM of excitons
at $T_{ex}=$ 90 K and $T_L=$ 5 K.
In the upper plot the calculations with
$\Lambda$ = 5, 10, and 20 nm are compared. ($\Delta$ is fixed at
one monolayer, 0.283 nm)
The diffusion coefficient obtained by fitting to the
simple diffusion model at the initial stage (0 - 25 ps) is
100, 170, and 250 cm$^2$ s$^{-1}$ for $\Lambda$ = 5, 10, and 20 nm,
respectively.
The lower plot shows the $\Delta$ dependence
(one monolayer; 0.283 nm, and two-monolayer; 0.566 nm)
of the transport when $\Lambda$ is fixed at 10 nm.
The initial diffusion coefficient is 170 an 60 cm$^2$ s$^{-1}$,
respectively.
These results show that the transport strongly depends on the
IFR parameters (the spatial spread is faster when the
lateral terrace size is larger and the terrace height is lower),
and thus the transport properties would vary from sample to sample
depending on the growth conditions.

We shall discuss the limitations of the present model with regard
to the realistic situation in the experiment.
The experimentally observed diffusion coefficients vary widely
from few cm$^2$ s$^{-1}$ to several hundreds of cm$^2$ s$^{-1}$
depending on the experimental conditions as sample temperatures,
excitation energies, carrier densities, and the sample
structures.
\cite{Hegarty,Ober,Smith,Yoon,Hill1,Hill2,Hill3,Tak1,Tak2,Akiyama}
The direct comparison of the present simulations with the
experimental observation is difficult
because, in experiments, free electrons and heavy holes are initially
photogenerated, then they form bound system, excitons.
Thus we should have started simulations from free $e$/hh's
and handled the exciton formation process properly.
However, though there are several experimental studies on the
exciton formation time\cite{Damen,Strobel,Blom,Amand}, to the best of
our knowledge, the theoretical formation model is not available.
Here we can compare the simulation results with experiments with
the similar carrier density conditions only
in qualitative manner, assuming that if the initial excitation
energy is higher, the resultant exciton temperature is also high.
In Ref. \onlinecite{Yoon,Tak1},
the experimental diffusivity decreases
with the excitation energy. The trend
agrees with the present simulations.
As for the sample temperature dependence,
\cite{Ober,Yoon,Hill1,Hill3,Tak1}
the experimental diffusivity decreases with sample temperatures
down to 20 K,
which also agrees with the simulation results if we assume that the
exciton temperature becomes lower with the sample temperatures.
However, below 20 K, Ref. \onlinecite{Hill3,Tak1} report the
increase of the diffusivity, which is not observed in the
other experiments.
Our simulation results does not reproduce the diffusivity
enhancement below 20 K.
The origin of the enhancement is not understood yet.

For the case of very lower exciton energy, we have to consider the
validity of the present IFR scattering model.
In the ex-IFR scattering an exciton is treated in a plane
wave mode, extending over all 2D plane. However it is experimentally
observed\cite{Christen,Zachau}
that excitons become localized in the potential minima of
the interface roughness in 2D plane with the reduction of
the in-plane kinetic energy.
The localized excitons can still migrate in the 2D plane by
phonon-assisted hopping or variable range hopping (in the very
low temperatures). This causes the significant
slowing down of the diffusivity\cite{Tak4}.
The present model does not include this effect.
This becomes important when the exciton temperature gets
sufficiently low by emitting phonons.

\section{Conclusions}
The time dependence of the spatial distributions of
nonequilibrium excitons was obtained using the
ensemble Monte Carlo simulations including
the interparticle scatterings (ex-ex, ex-carrier
carrier-carrier), LA phonon emissions/absorptions, and
the IFR scatterings.
The simulations show that the dominant scattering process
for excitons is
the ex-IFR scattering,
in contrast to charged carriers for which the carrier-carrier
scattering is most significant.
The difference of the dominant scattering
processes affects the energy relaxations: The $e$/hh system
approaches the quasi-thermal equilibrium among carriers very rapidly
(in few picoseconds)
through the efficient energy redistribution caused by the
carrier-carrier collisions. In contrast,
the excitons relax very slowly because their main scattering process,
the ex-IFR scattering, is elastic and does not change exciton
energies. Excitons can exchange energies
only through the ex-ex, ex-hh and ex-LA phonon
scatterings with very low probabilities.

The transport of excitons is essentially determined by their
average velocity
because the IFR scattering is weakly dependent on
the exciton distributions and the effect of other scattering
processes is negligibly small.
Thus the exciton transport can be regarded a diffusive
motion whose diffusion coefficient varies with time.
The diffusion coefficient, which is proportional to the
average velocity, varies through the energy exchange with
residual hh's and lattice.
The spatial transport is strongly dependent on the IFR parameters;
the lateral terrace size $\Lambda$ and the terrace height $\Delta$.
The transport is quenched when the terrace size is smaller or
the terrace height is higher.
This means that the exciton transport properties are sensitive to
the interface structures, and thus, to the growth conditions of
QW's.

\acknowledgments
We acknowledge Koji~Muraki for valuable discussions.
We also acknowledge Makoto~Takahashi for his offer
of the computer resources.
This work was partly supported by the Grant-in-aid for
Scientific Research No. 07855002 from the Ministry of
Education, Science and Culture, Japan.


\begin{figure}
\caption{The diagrams calculated in the exciton-electron ((a), (b))
and in the exciton-exciton ((c)-(f)) scatterings.}
\label{diagrams}
\end{figure}

\begin{figure}
\caption{The comparison of electron-heavy hole, exciton-heavy hole,
and exciton-exciton scattering rates plotted against momentum
transfer $q$.}
\label{compscatrate}
\end{figure}

\begin{figure}
\caption{Schematic diagram of the simulation area.
The initial excitons are generated
in the Gaussian shape with FWHM 6 $\mu$m.
They gradually spread out to the
surrounding area with the density distribution described by
the solid line.
The simulation area (radius 10 $\mu$m) is divided into concentric
circular regions with 1 $\mu$m step, and the average carrier density
at each region is used in the evaluation of carrier scatterings. }
\label{simarea}
\end{figure}

\begin{figure}
\caption{The numbers of scattering events per particle per second
for electrons (upper) and excitons (lower) with the initial
excitation energy 10 meV and at 5 K.
Notice that in the lower case the data are plotted in
the logarithmic scale.}
\label{No.ofScat}
\end{figure}

\begin{figure}
\caption{The time evolution of the momentum distribution functions
for (a) electrons and (b) excitons, from 0
ps (near side) - 15 ps (far side) with the time step
1 ps. The initial excitation
energy is 10 meV and the lattice temperature is 5 K.}
\label{energyrelaxation}
\end{figure}

\begin{figure}
\caption{The momentum distribution functions of excitons
from 0 ps (near side) to 500 ps (far side) with 25 ps step.
The initial momentum distribution is in the Bose function
with $T_{ex} =$ 30 K and the lattice temperature at 5 K.}
\label{excitonrelax}
\end{figure}

\begin{figure}
\caption{The FWHM of the exciton spatial distributions from 0 to
500 ps with the initial exciton temperatures changed as a
parameter. The lattice
temperature is fixed at 5 K. The FWHM is obtained by fitting the
Gaussian function to the spatial exciton distributions calculated
in the simulations.
The FWHM of the simple diffusion model with $D =$ 30 cm$^2$ s$^{-1}$
is also plotted for reference.}
\label{Fig7FWHM}
\end{figure}

\begin{figure}
\caption{The FWHM of the exciton spatial distributions at
the lattice temperature 30 K with $T_{ex} =$ 30 K (upper) and
5 K (lower).}
\label{Fig8FWHM}
\end{figure}

\begin{figure}
\caption{The FWHM of the exciton spatial distributions at
different interface roughness parameters: the correlation length
$\Lambda$ dependence (upper) and the width fluctuation $\Delta$
dependence (lower).}
\label{Fig9IFR}
\end{figure}

\end{document}